\documentclass{elsarticle}

\usepackage{hyperref,amssymb,amsmath}

\usepackage[latin1]{inputenc}
\usepackage{graphicx}
\usepackage{xcolor,tikz}
\usepackage{pgfplots}
\pgfplotsset{width=10cm,compat=1.9}
\usepackage{algorithm}
\usepackage{algorithmic}
\usepackage{graphicx}
\usepackage{subcaption}
\usepackage{soul}

\usepackage{algorithm}

	\definecolor{naranjauca}{cmyk}{ 0, 0.6, 1, 0}

\newdefinition{definition}[theorem]{Definition }
\newdefinition{remark}[theorem]{Remark }
\newdefinition{example}[theorem]{Example }
\newproof{proof}{Proof}

\journal{International Journal of Information Technology \& Decision Making}

\makeatletter
\def\ps@pprintTitle{%
  \let\@oddhead\@empty
  \let\@evenhead\@empty
  \let\@evenfoot\@oddfoot
}
\makeatother

\begin{document}

%
%
%

\begin{frontmatter}

\title{
Decision support system for photovoltaic fault detection avoiding meteorological conditions}

\author[uc]{Roberto G. Aragón\corref{co}}
\ead{roberto.aragon@uca.es}

\author[uc]{M. Eugenia Cornejo}
\ead{mariaeugenia.cornejo@uca.es}

\author[uc]{Jes{\'u}s Medina}
\ead{jesus.medina@uca.es}

\author[uclm]{Juan Moreno-Garc\'ia}
\ead{Juan.Moreno@uclm.es}

\author[uc]{Eloísa Ramírez-Poussa}
\ead{eloisa.ramirez@uca.es}

\address[uc]{Department of Mathematics, University of Cádiz, Cádiz, Spain}
\address[uclm]{Escuela de Ingenier\'ia Industrial, Universidad de Castilla La Mancha,  Toledo, Spain}

\begin{abstract}

{
A fundamental issue about installation of photovoltaic solar power stations is the optimization of the energy generation and the fault detection, for which different techniques and methodologies have already been developed considering meteorological conditions. This fact implies the use of unstable and difficult predictable variables which may give rise to a possible problem for the plausibility of the proposed techniques and methodologies in particular conditions. In this line, our goal is to provide a decision support system for photovoltaic fault detection avoiding meteorological conditions.

This paper has developed a mathematical mechanism based on fuzzy sets in order to optimize the energy production in the photovoltaic facilities,  detecting anomalous behaviors in the energy generated by the facilities over time. Specifically, the incorrect and correct 
 behaviors of the photovoltaic facilities have been modeled through the use of different membership mappings. From these mappings, a decision support system based on OWA operators informs of the performances of the facilities per day, by using natural language. Moreover, a state machine is also designed to determine the stage of each facility based on the stages and the performances from previous days.
The main advantage of the designed system is that it solves the problem of ``constant loss of energy production'', without the consideration of meteorological conditions and  being able to be more profitable. Moreover, the system is also scalable and portable, and complements previous works in energy production optimization. Finally, the proposed mechanism has been tested with real data, provided by Grupo Energético de Puerto Real S.A. which is an enterprise in charge of the management of six photovoltaic facilities in Puerto Real, Cádiz, Spain, and good results have been obtained for faulting detection. }

\end{abstract}


\end{frontmatter}


\section{Introduction}\label{sec:introduction}

One important part of the Europe 2020 strategy is focused on reaching 20\% of gross final energy consumption from renewable sources by 2020, and {increasing this trend to reach} at least 32\% by 2030. This goal is also pursued by other countries and continents, which also try to fostering the implementation of renewable energy. As a consequence, many research institutions are developing new mechanisms in different  research lines, such as in new construction protocols, management, legal aspects, cybersecurity, and the application of artificial intelligence tools~\cite{chang2014,heldeweg2020,kobylinski2020,leszczyna2019,morato2020,review6}. This last line   mainly tries to enhance the behavior of the facilities and optimizing the energy generation~\cite{marin2019,mattiussi2014,sandeep2019,singh2019,torreglosa2016}.   

In the framework of renewable energy,  we have many sources, such as biofuel, geothermal,  hydroelectricity, tidal, solar, wave, and  wind, being solar energy one of the most important. Many companies around the {world} are focused on the design of photovoltaic (PV) cells, required accessories, and in the installation of photovoltaic solar power stations. Hence, the innovation of more efficient generation of photovoltaic power is a fundamental goal~\cite{BENAMMAR20201016,SAID20201274}. 
In this aspect, fault detection is a critical issue for   optimizing the energy generation and encouraging the use of renewable energy.  In this research line, many groups are working {and different mechanisms have been proposed, as the authors in~\cite{kordestani18} also highlighted. They for example  apply}  (artificial) neural networks, regression, multi-gene genetic programming, multi-objective and multi-attribute decision-making, and principal component analysis, among others~\cite{chenIVcharact2017,dhimishsixlayer2017,dhimishPVCurves2017,hajjiMultivariate2020,hussainANN2020,khalilComparative2020,manohar2020,rodriguesMLPV2020,samara2020,takashima2009}. 
{One important fact, also noted in~\cite{chenIVcharact2017}, is that fault diagnosis models are not efficient to be trained and updated, because usually the  PV arrays' operating point only contains little information. In the aforementioned paper, the emerging kernel based extreme learning
machine (KELM),  the Nelder-Mead Simplex
(NMS) optimization method and a Simulink based PV modeling approach are combined to present a fault diagnosis model. However, the final result is a complex system which considers variables directly dependent on the operating
condition, i.e., the input irradiance and operating temperature.
Moreover, they required restrictive boundary conditions, such as all the solar cells in the PV module be homogeneous.

The authors in 
\cite{dhimishPVCurves2017} have proposed a smart algorithm based on support vector machine (SVM) technique for diagnosis short circuit faults in PV systems,  which requieres irradiance level and PV module's temperature. \cite{dhimishsixlayer2017} also  analyzes  the theoretical voltage ratio (VR) and power ratio (PR) of the 
the given  grid-connected photovoltaic (GCPV) plant, but  irradiance level and PV module's temperature are also required. 

In~\cite{hajjiMultivariate2020},
the principal component analysis (PCA) technique is used for extracting and selecting the most relevant multivariate features, and the supervised machine learning (SML) classifiers are applied to faults diagnosis. It also considers   outdoor conditions, such as  temperature, irradiation and humidity. In addition, it is only a theoretical method designed on  an emulator of a GCPV system.

The method proposed in 
\cite{hussainANN2020} is based on artificial neural networks
(ANN) and considers a radial basis function (RBF), requiring only
two parameters as the input to the ANN (solar irradiance and output power).  The authors highlight that the fault detection accuracy strongly depends on the varying nature of each PV plant. 

The same input variables are also taken into account in~\cite{rodriguesMLPV2020}. The authors introduce a hybrid methodology which combines the use of a fuzzy system and the use of a machine learning system containing five different trained machine learning models, such as the regression tree, artificial neural networks, multi-gene genetic programming, Gaussian process, and support vector machines for regression. The Takagi-Sugeno type fuzzy logic inference system is used to select one of the five machine learning techniques that would provide the most accurate prediction, by analyzing the solar irradiation input that is given.
	
In~\cite{khalilComparative2020}, the mathematical formulation of the different  processed carry out in the energy generation is considered, such as the current and voltage characteristics of the solar cell, which also require of the irradiance
and temperature of PV module, among others.

\cite{samara2020} highlights the required cost and complex hardware required by the majority of the current fault diagnostic techniques. The authors present in the aforementioned paper a fault diagnosis algorithm based on an artificial intelligent nonlinear autoregressive exogenous (NARX) neural network and Sugeno fuzzy inference.
The fuzzy inference requires the actual sensed PV system output
power, the predicted PV system output power, and the sensed surrounding conditions. Hence, this method also needs meteorological values and   highly monitored plants.

Probability theory, discrete wavelet transform (DWT) and an optimal random forest (ORF) based classifier are combined in~\cite{manohar2020} to design a procedure to fault detection/classification and zone identification, which is also meteorological data-dependent.
}

{The main problem of the previous mentioned methodologies is that they consider meteorological conditions. Notice that, for example, PV system energy production is non-linear because it is influenced by the
random nature of weather conditions.}
 As a consequence, different difficult predictable and  unstable variables are taken into account in the different developed methodologies, which shows a possible problem for the plausibility of the introduced methodologies in particular conditions. For example, the strong winds  in Cádiz, Spain, remarkably decrease the accuracy meteorological predictions and the meteorological stations must be protected from them, which also decreases their accuracy. {Other authors have considered this challenge previously, however, the previous restrictions are very strong to be realistic and portable. For example, the methodology proposed in~\cite{takashima2009} does not depend on the meteorological conditions, but it needs strong requirements, such as every string must be monitored and  the method must be applied to identical strings, which is complicated in usual facilities. Maybe, this method can work in some new installations, but over time different components (investors, photovoltaic cells, etc.) must be replaced by new ones with different features and performance.}

On the other hand, the {daily} maintenance of (medium and small)  photovoltaic solar power stations can be very expensive. Therefore, the design of a decision support systems for unsupervised stations is another required challenge. The creation of an automatic intelligence system becomes even more useful when the system provides the outputs in natural language easily understandable and executable by the station manager.

This paper will introduce a mathematical strategy based on fuzzy sets~\cite{Z1} with the purpose of improving the processes involved in the management of renewable energies and energy efficiency, avoiding the consideration of surrounding conditions, unlike other methodologies~
\cite{dhimishsixlayer2017,dhimishPVCurves2017,rodriguesMLPV2020,samara2020}. We will show that the introduced strategy has good results for faulting detection. Specifically, the company Grupo Energ\'etico de Puerto Real S.A. (GEN), located in Puerto Real, C\'adiz (Spain), is commended with the management and integral maintenance of the public lighting network of the city. Its main purpose is guaranteeing a service to the citizen of the highest quality, minimizing operating costs and impacts to the environment, through energy efficiency.  
{Hence,} one of the activities of GEN is the generation of energy through six {heterogeneous} photovoltaic facilities. {Take into account that these plants have different number and brands of photovoltaic cells, investors, etc. Indeed, they started in different periods.}  

{Therefore, o}ne of the most important and critical aspects related to manage and control this process is the associated degree of imprecision, incompleteness, uncertainty and unpredictability, which highlights the need of using mathematical fuzzy techniques.

 Specifically, a decision support system has been presented in this paper to detect anomalous behaviors in the energy generated by the facilities over time. This system has been created by using a mathematical mechanism to compare the energy production among sections (substations) of a big facility or different close facilities, which can be scalable and portable {to small, medium and big PV plants. It is clear that a big plant needs to be divided into small subplants for a better monitoring, and also a small plant could also be split. For example, if the plant is installed in a common area of three different neighborhood communities (with different number of people) for self-sufficiency, the installation must be divided into three non-proportional parts, although the PV plant could be relatively small.} The main contributions presented in this {paper} are listed below:

 \begin{itemize}
	\item From the available information about the days in which the facilities have had  correct or  incorrect performances, we have designed different functions to describe the behaviors of the photovoltaic facilities.
	
	\item Taking into account the previously mentioned functions, we have created a decision support system to inform of the performances of the facilities per day, by using natural language. 
	
	\item We have designed a {state machine} that allows us to determine the stage of each facility, according to the stages and the performances obtained in the previous days for each facility.
	
	\item {The designed system has been tested on real photovoltaic stations  during {10} months obtaining good results, which have been analyzed in the last section of the paper.} 

 \end{itemize}
 
This system is being very useful and it is in production in GEN. Before this system, this company detected the days when any of the stations had an abnormal behavior by making a visual comparison of the daily graphs of the generation of each facility. This caused a waste of time for the person in charge of this task.  Since this person was in charge of other functions in the company, many times the data series were not reviewed for several days. This fact caused an accumulation of the error over time. As a result, economic losses occurred since the facilities generate less energy than it is expected. For that reason, the company was interested in the introduced system to receive daily reports related to the status of the facilities, in a completely automatic way and using natural language.

{The paper is organized as follows. Section~\ref{sec:background} briefly reviews the dataset considered in this study and recalls the definition and some properties of ordered weighted averaging aggregation operators. Section~\ref{sec:comparative} explains the proposed mathematical reasoning to compare the energy production among the photovoltaic facilities, which gives rise to a decision support  system to detect anomalous behaviors in the energy generated by the facilities over time. Section~\ref{sec:example} is devoted to illustrate and analyze the results obtained by the automatic system described in Section~\ref{sec:comparative}. Finally, Section~\ref{sec:conclusions} finishes with some conclusions and prospects for future work.}

\section{{Preliminaries}}\label{sec:background}

GEN supplies a dataset of a multivalued time series that has the information of the energy generated by each station along with the date and time.
{Each station has different number of photovoltaic cells and investors, which implies that each station has different peak power\footnote{It is the maximum value of energy that the photovoltaic cells can  generate per facility, due to its technical specifications} and rated power\footnote{It is the maximum value of energy that the investors of each facility convert from direct current (DC) to alternating current (AC), due to its technical specifications}. For example, the peak power ranges from 47,61Kw to  91.125Kw.}
These data  are preprocessed, sorted by date, time and station, in order to be incorporated into the database on which we will carry out the study.

From the collected daily information,  a mathematical  mechanism to compare the energy production is proposed. This mechanism consists of several phases which will be detailed in the following section. One of these phases {makes} use of a special kind of mathematical aggregation operator called ordered weighted averaging (OWA), {which}  were introduced by Yager in the eighties~\cite{yager88} as a solution to the concerned problem of aggregating  multi-criteria to provide an overall decision making. {On the other hand, in order to analyze the outputs obtained by the decision system we will make use of  confusion matrices~\cite{Berk2008}. Confusion matrices let us measure the accuracy and error obtained in the predictions provided by the decision system in the test done. In our case, they will allow to evaluate the classification of states of each facility. For these reasons, in the following, we will briefly recall some preliminaries about these tools.

\subsection{OWA operators}

OWA operators have been studied from a theoretical~\cite{Beliakov2020,demiguel2017,LIZASOAIN2013,MEDINA202138,PU201924,YAGER201788,yagerkacprzyk97,review1,review3,review5} and applied point of view~\cite{review2, review4}. 
For example, they have been defined on complete lattices~\cite{medina2013,Medina12Charac,mesiar08}, in order to allow the  consideration of incomparable elements to be aggregated~\cite{demiguel2017,LIZASOAIN2013}. 
OWA operators have been applied to many fields, such as in economy, 
linguistic, clustering, energy, urban wastewater, etc. For example, 
a {multi-attribute} decision-making method was introduced in~\cite{garg2019} to deal with linguistic membership and nonmembership degrees. They have also been applied to linguistic decision judgments~\cite{CHIAO2019}, noise removal in computer vision~\cite{JAIME201464}, 
in fuzzy clustering to handle noise and outliers~\cite{SIMINSKI2017}, 
in order to
neutralize the negative effects of possible outlier fuzzy data in the clustering process~\cite{DURSO2019}, to  provide a decomposition for all the rank-dependent poverty measures in terms of incidence, intensity and inequality~\cite{ARISTONDO2016},
to allow experts to express multiple self-confidence levels when providing their preferences~\cite{xia2019}, etc.

Specifically, an \emph{$n$-ary ordered weighted averaging} (OWA) operator is a function $F_W\colon [0,1]^n\to [0,1]$   associated with  a list of  weights $W=\{w_1,\dots,w_n\}$, such that $w_i\in [0,1]$ and $\sum_{i=1}^n w_i=1$, and defined as 
$$
F_W(a_1,\dots,a_n)= \sum_{i=1}^n w_i a_{\rho (i)}
$$ 
where $\rho\colon \{1,\dots, n\}\to \{1,\dots, n\}$ is a permutation on the index set satisfying that 
$ a_{\rho(n)}\leq\dots \leq a_{\rho(1)} $.
Hence, the greatest value of the tuple $(a_1,\dots,a_n)$ is multiplied by the first weight $w_1$, the second to the second weight and so on. Since two values  $a_i$ and $a_j$, with $i\neq j$, can be equals: $a_i=a_j$, at least another different permutation $\sigma$ {can}  exist satisfying the ordering among the values: $ a_{\sigma(n)}\leq\dots \leq a_{\sigma(1)} $. However, the function $F_W$ is well defined because, if this happens, the equality  $\sum_{i=1}^n w_i a_{\rho (i)}=\sum_{i=1}^n w_i a_{\sigma (i)}$ holds. Since $n=1$ provides a trivial case, usually  $2\leq n$, as we will consider in the proposed mechanism.

The mathematical formulation of OWA operators allows the user to consider multicriteria, from the least case given by the conjunction ``and'' and represented by the minimum operator, and the greatest one given by the disjunction ``or'' and represented by the maximum. 
  Therefore, 
  $$\min\{a_1,\dots,a_n\}\leq F_W(a_1,\dots,a_n)\leq \max\{a_1,\dots,a_n\}
  $$ for all  $(a_1,\dots,a_n)\in [0,1]^n$.
  Moreover, two other very interesting properties of the OWA operators are: 
  \begin{itemize}
\item  {Idempotency}: $F_W(a,\dots,a)= a$, for all $a\in [0,1]$.
\item  {Symmetry}:  $F_W(a_1,\dots,a_n)=F_W(a_{\sigma (1)},\dots,a_{\sigma (n)})$, for all permutation $\sigma\colon \{1,\dots, n\}\to \{1,\dots, n\}$ and $(a_1,\dots,a_n)\in [0,1]^n$.
\end{itemize}

\subsection{Confusion matrix}

The confusion matrix is a well-known tool for assessing how good a classification model is [4]. In particular, it serves to show when one class is confused with another one. In the confusion matrix, the rows correspond to predicted conditions and the columns corresponds to actual conditions. Therefore, this kind of matrices are suitable to detect false positives (FP), false negatives (FN), true positives (TP), true negatives (TN), as well as the errors obtained in the predictions. Each of these cases indicates the following: TP for correctly automatic detection alert values, FP for incorrect automatic detection alert values, TN for correctly automatic detection no-alert values, FN for incorrect automatic detection no-alert values. In addition, the errors are computed as Table~\ref{confusion_matrix} shows.

\begin{table}[!h]
	\centering
	\renewcommand{\arraystretch}{1.5}
	\begin{tabular}{|c|c|c|c|}
		\hline
		& No Alert & Alert &  Model error \\ \hline
		No Alert	 & TN       & FN 	& 	$\frac{\text{FN}}{\text{FN}+\text{TN}}$	\\ \hline
		Alert		 & FP       & TP	& 	$\frac{\text{FP}}{\text{FP}+\text{TP}}$	\\ \hline
		Error of use & 	$\frac{\text{FP}}{\text{TN}+\text{FP}}$		&  $\frac{\text{FN}}{\text{FN}+\text{TP}}$		& $\frac{\text{FN} + \text{FP}}{\text{TN}+\text{FN}+\text{FP}+\text{TP}}$		\\ \hline
	\end{tabular}
	\caption{Confusion matrix model.}
	\label{confusion_matrix}
\end{table}
}

\section{
Decision support system to  detect and alert of abnormal performances 
}\label{sec:comparative}

In this section, we will describe a decision support system to detect anomalous behaviors in the energy generated by the facilities. The main idea in which the proposed automatic detection system is based on is the following: since all the facilities are located in the same city and, as a consequence, they are under the same climatic conditions,  it is expected that, when the facilities work in a suitable way, the  differences of the energy generation between one facility and another one should be around a constant value. That is, the difference between the production of two facilities must always oscillate in a certain range of values, as long as the facilities are working properly. Consequently, when this difference is out of the expected range of values, the automatic decision system will indicate an anomalous behavior in the production of one of the photovoltaic facilities considered in the comparison. Moreover, this methodology can be exportable and scalable to photovoltaic energy stations with investors, splitting the collection of energy in different substations and comparing them using the mechanism introduced in this paper.

The mentioned decision support system is supported by a theoretical reasoning which allows us to compare the energy produced by the different photovoltaic facilities in order to detect possible reductions in the performance of the facilities. To explain this theoretical reasoning, we will establish some notation conventions. The company has six different photovoltaic facilities, we will use the notation $\mathcal{I}_i$ in order to refer to the photovoltaic facility $i$, where $i\in\{1,\dots, 6\}$. 
We will use the notation $GE_{ij}$ to refer to the generated energy by the photovoltaic facility $\mathcal{I}_i$ in the hour $j$, with $j\in \{0,\dots,23\}$, {these values are measured and provided by the enterprise}. Moreover, for testing, we have a learning dataset on days in which the facilities have had a correct performance and days in which they have had an incorrect performance. We will take advantage of the information available regarding the correct and incorrect days of each facility to create the decision system. 
 {Specifically, a learning process  that will allow us to classify the daily behavior of each facility, based on the information contained in that learning dataset, will be established (the learning procedure will be detailed in Section \ref{sec:learning}). As a result, we will able to detect those days in which a particular facility has  an abnormal performance (the procedure of detection will be explained in Section \ref{sec:detect_anomalous_behavior}).}

\subsection{{Learning to classify the performance of the facilities}\label{sec:learning}}
This section will show the procedure to classify the performance of each facility. In particular, {taking into account the GEN dataset of 2019,} different parameters will be studied in order to be able to correctly classify the daily performance of the facilities.

Specifically, the decision support system will calculate the difference of the energy production for each pair of photovoltaic facilities basing on the previously mentioned idea. These differences will be evaluated in membership functions that will catalog the energy production of the facilities as suitable or abnormal. This procedure is explained in a detailed way in Algorithm \ref{alg:learning_membership_functions} that uses as input the learning dataset. The steps carried out by the algorithm are detailed next. 

\begin{algorithm}
	\caption{{Learning daily performance parameters}}
	\label{alg:learning_membership_functions}
	\begin{algorithmic}[1]
		\STATE Calculate the total performance of energy production by each  photovoltaic facility.
		\STATE Compute the difference of the total performances of energy production (Daily Performance Matrix $DPM$).	
		\FOR{each facility $\mathcal{I}_i$ with $i\in\{1,\dots,6\}$}
		\STATE Divide the daily information in correct days $D_C^{i}$ and incorrect days $D_I^{i}$.
		\ENDFOR
		\STATE  {Compute the parameters that define the membership functions.}
		
	\end{algorithmic}
\end{algorithm}

\begin{description}
\item [Line 1:] First of all, from the information of the energy production per photovoltaic facility contained in our database, we calculate the total performance of daily energy production. The following formula computes the normalization of the total performance of energy production by the photovoltaic facility $\mathcal{I}_i$, with $i\in \{1,\dots, 6\}$, at day $d$:
$$
\rho_i(d)= \displaystyle\frac{\sum_{j=0}^{23}GE_{ij}(d)}{PP_{i}(d)}*100
$$
{Notice that, the normalization arises from the ratio of the generated energy {($GE_{ij}(d)$)} and peak power {($PP_{i}(d)$)} per day $d$}. It is also important to mention that, in the previous formula, we consider the peak power, instead of the rated power, because the peak power is directly related to the photovoltaic cells that generate the energy. Moreover, the generated energy is always lesser than the peak power. 

\item [Line 2:]
Once we have the total performances of energy productions of the photovoltaic facilities, we can compute the difference between the total performances of energy production provided by two photovoltaic facilities, $\mathcal{I}_i$ and $\mathcal{I}_k$ with $i,k\in\{1,\dots,6\}$,  at day $d$, by means of the following formula:
$$
\delta_{ik}(d)=\displaystyle\frac{\rho_i(d)-\rho_k(d)}{\max\{\rho_i(d), \rho_k(d)\}}*100
$$
It is important to mention that in the previous formula the difference $\rho_i(d)-\rho_k(d)$ is divided by $\max\{\rho_i(d), \rho_k(d)\}$ in order to obtain a relative difference, which provides an easier interpretation of the result as well as an easier way to establish comparisons between two facilities $\mathcal{I}_i$ and $\mathcal{I}_k$ over time (for a fixed period of days).
{Moreover,  we have that  $\delta_{ik}=-\delta_{ki}$. Notice that the sign is important in order to know what station usually generates more energy than the others.}  {All this information is stored in a daily matrix called ``daily performance matrix''.} 
\item [Lines 3-5:] 
Furthermore, for each facility $\mathcal{I}_i$ with $i\in\{1,\dots,6\}$, we will separate the information contained in the database in two parts attending to the kind of day, correct and incorrect denoted as $D_C^{i}$ and $D_I^{i}$ respectively. 
We can have days with an unclear classification, which will be removed for the test. 
{Anyway, we already have a lot of noise in the data, due to the non-daily manual classification of correct and incorrect days. This is an important problem to take into account and that the use of fuzzy sets will remarkably reduce its impact in the final result.}

\item [Line 6:] The main idea in which this step is based on is that, from the correct days, we can know what can be considered an acceptable difference and, from the incorrect days, we can detect anomalous differences. In order to simplify the explanation of this step,  we will focus the attention on the analysis of a specific photovoltaic facility $\mathcal{I}_i$.
{For each $\mathcal{I}_k$, with $k\neq i$}, we select those days $d$ in which {both} facilities have had a correct performance and we compute the relative differences $\delta_{ik}(d)$.
 The least value reached, denoted as $b_{ik}$, will be the least value indicating a suitable performance of the facility $\mathcal{I}_i$ with respect to $\mathcal{I}_k$, with $k\neq i$. 
When the number of days in which both facilities have had a correct performance is equal to zero, that is $D_C^{i}\cap D_C^{k}=\varnothing$, then the value $b_{ik}$ will be identified from the value $a_{ik}$, whose computation is explained in the following paragraph.

Considering the information of the days in which the fixed photovoltaic facility $\mathcal{I}_i$ has an incorrect behavior and the rest of  facilities $\mathcal{I}_k$, with $k\neq i$, have a correct behavior, we compute the daily relative differences of performance. Under these conditions, the greatest value of $\delta_{ik}(d)$, for each $k\neq i$, denoted as $a_{ik}$, will determine an interval $[a_{ik},b_{ik}]$ indicating an anomalous performance of the facility $\mathcal{I}_i$ with respect to $\mathcal{I}_k$ with $k\neq i$. Obviously, any relative difference lower than $a_{ik}$ will indicate a bad performance of the facility $\mathcal{I}_i$ with respect to $\mathcal{I}_k$. 
Formally, the values $a_{ik}$ and $b_{ik}$, for each $k\neq i$, will be obtained by means of the following expressions:
\begin{eqnarray*}
a_{ik}&=&\max\{\delta_{ik}(d)\mid d\in D_I^{i}\cap D_C^{k}\}\\
b_{ik}&=&\min\{\delta_{ik}(d)\mid d\in D_C^{i}\cap D_C^{k}\}
\end{eqnarray*}

It must be verified that $a_ {ik}$ is lesser than $ b_ {ik}$. Otherwise, an inconsistency in the data will be detected{, which is produced by an inaccurate classification of correct and incorrect days of each facility}. In such a case, the roles of the values $a_ {ik}$ and $b_ {ik}$ will be exchanged, that is:
\begin{eqnarray*}
a_{ik}&=&\min\{\delta_{ik}(d)\mid d\in D_C^{i}\cap D_C^{k}\}\\
b_{ik}&=&\max\{\delta_{ik}(d)\mid d\in D_I^{i}\cap D_C^{k}\}
\end{eqnarray*}
\end{description}
 
 For instance, if we focus on the facility $\mathcal{I}_1$ and choose the facility $\mathcal{I}_2$ to compare with it. In the computation of the interval $[a_{1 2}, b_{1 2}]$, we obtain the values $a_{1 2} = -12.550$ and $b_{1 2} = -21.662$. As we can observe $a_{1 2} > b_{1 2}$, hence we have to exchanged these values, obtaining the final interval:  $[-21.662, -12.550]$. Table~\ref{inconsistencias} shows a comparative matrix between facilities indicating with   1 that the role of values $a_{i k}$ and $b_{i k}$ have been exchanged and 0, otherwise.
\begin{table}[h!]
\begin{center}
\begin{tabular}{|c|cccccc|}
\hline
& $\mathcal {I}_1$ & $\mathcal {I}_2$ & $\mathcal {I}_3$ & $\mathcal {I}_4$ & $\mathcal {I}_5$ & $\mathcal {I}_6$\\ \hline
$\mathcal {I}_1$ & $-$ & $1 $ & $1 $ & $1 $ & $1 $ & $1 $ \\ \hline
$\mathcal {I}_2$ & $0 $ & $-$ & $0$ & $0 $ & $0 $ & $0 $ \\ \hline
$\mathcal {I}_3$ & $0 $ & $1 $ & $-$ & $1 $ & $1 $ & $1 $ \\ \hline
$\mathcal {I}_4$ & $0 $ & $0 $ & $1 $ & $-$ & $0 $ & $0 $ \\ \hline
$\mathcal {I}_5$ & $0 $ & $1 $ & $0 $ & $1 $ & $-$ & $1 $ \\ \hline
$\mathcal {I}_6$ & $1 $ & $1 $ & $1 $ & $1 $ & $1 $ & $-$ \\ \hline
\end{tabular}
\caption{Table of comparative for each pair of photovoltaic facilities that indicates if an inconsistency in the data has been detected.}\label{inconsistencias}
\end{center}
\end{table}

\begin{figure}[h!]
\begin{minipage}{.55\textwidth}
$$
\mu_{ik}(x) = \left\{
\begin{array}{ll}
0      & \mathrm{if\ } x \leq a_{ik} \\
\frac{x-a_{ik}}{b_{ik}-a_{ik}} & \mathrm{if\ } a_{ik}< x< b_{ik}\\
1 & \mathrm{if\ }  b_{ik}\leq x\\
\end{array}
\right.
$$
\end{minipage}
\begin{minipage}{.25\textwidth}
\begin{center}
\includegraphics[width=5.2cm]{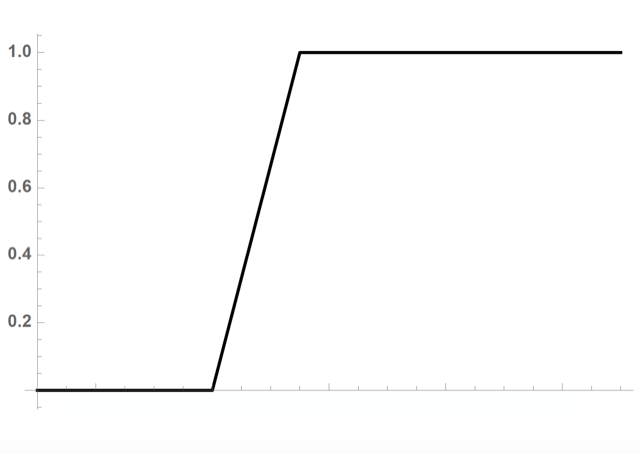}
\end{center}
\end{minipage}
\caption{Membership function indicating the degree of suitable performance of the facility $\mathcal{I}_i$ with respect to $\mathcal{I}_k$.}\label{fig:trapecio}
\end{figure}

We will build a trapezoidal membership function, from the values $a_{ik}$ and $b_{ik}$ for each $k\neq i$, to evaluate the daily relative differences of performance of the photovoltaic facilities $\mathcal{I}_i$ and $\mathcal{I}_k$. These membership functions $\mu_{ik}\colon\mathbb{R}\rightarrow [0,1]$ are defined as Figure~\ref{fig:trapecio} shows and they will be used to indicate the degree of suitable performance of the facility $\mathcal{I}_i$ with respect to $\mathcal{I}_k$. Hence, each facility will be associated with five different membership functions. 

\begin{figure}[h!]
\begin{minipage}{.55\textwidth}
$$
\mu_{ik}(x) = \left\{
\begin{array}{ll}
0      & \mathrm{if\ } x < b_{ik} \\
1 & \mathrm{if\ } x \geq b_{ik}\\
\end{array}
\right.
$$
\end{minipage}
\begin{minipage}{.25\textwidth}
\begin{center}
\includegraphics[width=5.2cm]{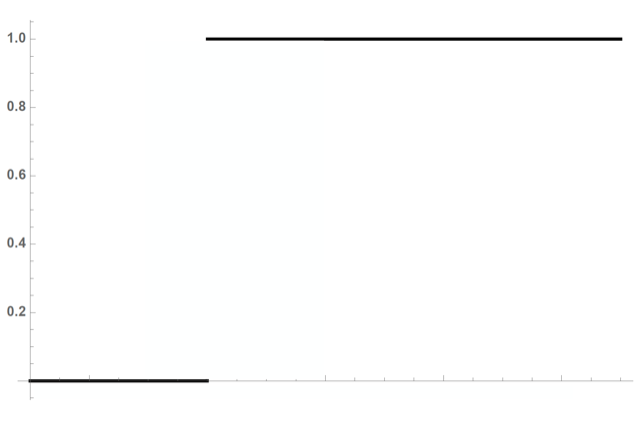}
\end{center}
\end{minipage}
\caption{Membership function indicating the degree of
suitable performance of the facility $\mathcal{I}_i$ with respect to $\mathcal{I}_k$, when we do not have information about the incorrect days neither $\mathcal{I}_i$ nor $\mathcal{I}_k$.}\label{fig:trapecio2}
\end{figure}

Note that, if $D_I^{i}\cap D_C^{k}=\varnothing$, then $a_{ik}$ will be zero and, as a consequence, the system  can provide bad classifications. This case arises for example when no information related to incorrect days for station $i$ is available.  Therefore, this case must be studied independently considering two cases. 
(1) If we have information about incorrect days of the facility $\mathcal{I}_k$, specifically $D_I^{k}\cap D_C^{i}\neq\varnothing$ holds, then we can determine the value $a_{ik}$ by symmetry, from the values $a_{ki}$, $b_{ki}$ and $b_{ik}$ by using  the formula $a_{ik}=b_{ik}-(b_{ki}-a_{ki})$. With this formula, we make that the distance between the values $a_{ik}$ and $b_{ik}$ coincides with the distance between $a_{ki}$ and $b_{ki}$, which has sense because the values $a_{ki}$ and $b_{ki}$ are the resulting values obtained from the computation of the opposite differences between the facilities $\mathcal{I}_i$ and $\mathcal{I}_k$.  
(2) If we do not have information about the incorrect days of  $\mathcal{I}_k$ for the correct days of $\mathcal{I}_i$ ($D_I^{k}\cap D_C^{i}=\varnothing$), then we will consider that $a_{ik}=b_{ik}$. This fact will cause that any relative difference lesser than the ones registered for the facilities $\mathcal{I}_i$ and $\mathcal{I}_k$ will produce an alert. In this case, the membership function will be defined as in Figure~\ref{fig:trapecio2}. {Fortunately, in the considered dataset this  second case does not needed to be considered.    Table~\ref{fig:symmetries} shows with 1 in which combinations $a_{ik}$ has been computed by  symmetry (case (1) detailed above).

\begin{table}[h!]
	\begin{center}
		\begin{tabular}{|c|cccccc|}
			\hline
			& $\mathcal {I}_1$ & $\mathcal {I}_2$  &  $\mathcal {I}_3$ &  $\mathcal {I}_4$ & $\mathcal {I}_5$ & $\mathcal {I}_6$\\ \hline
			$\mathcal {I}_1$ & $-$  & $0$  &  $0$   &  $0$   & $0$   & $0$  \\  \hline
			$\mathcal {I}_2$ & $1$  & $-$  &  $1$   &  $1$   & $1$   & $1$  \\  \hline
			$\mathcal {I}_3$ & $1$  & $0$  &  $-$   &  $0$   & $0$   & $0$  \\  \hline
			$\mathcal {I}_4$ & $1$  & $0$  &  $0$   &  $-$   & $0$   & $0$  \\  \hline
			$\mathcal {I}_5$ & $1$  & $0$  &  $0$   &  $0$   & $-$   & $0$  \\  \hline
			$\mathcal {I}_6$ & $0$  & $0$  &  $0$   &  $0$   & $0$   & $-$  \\  \hline
		\end{tabular}
		\caption{Table of comparative for each pair of photovoltaic facilities  indicating  if  $a_{ik}$ has been computed by  symmetry.}\label{fig:symmetries}
	\end{center}
\end{table}}

Repeating this process, we will obtain 5 membership functions to determine 5 degrees of proper performance of the facility $\mathcal{I}_i$ with respect to the rest of facilities.

{The methodology presented  in this section, including the membership functions that describe the behaviors of the facilities together with the
rest of introduced formulas, have been implemented in the automatic system.}

\subsection{{Detecting the abnormal performance of the facilities}\label{sec:detect_anomalous_behavior}}

This section will detail how the decision support system can detect anomalous performance of the facilities for a particular day $d^*${, and}  how the system classifies the performance over time by using a state machine    {as well.}

{
\begin{algorithm}
	\caption{{Detecting abnormal performance}} 
	\label{alg:detecting_abnormal_performance}
	\begin{algorithmic}[1]
		\STATE Calculate the total performance of energy production by each photovoltaic facility.
		\STATE Compute the difference of the total performances of energy production.	
		\STATE Compute the membership function for each difference from the matrix of evaluation of the relative differences.
		\STATE Apply OWA operator to each row.
		\STATE Obtain a linguistic description about the behavior of each facility.		
	\end{algorithmic}
\end{algorithm}

{{Algorithm~\ref{alg:detecting_abnormal_performance} describes the method used to detect the abnormal performance of the facilities. The steps to detect anomalous behaviors for a particular day $d^*$ are the following:}}
}
\begin{description}
\item [Line 1:] We compute the total performance of the relative energy production $\rho_i(d^*)$, for all the photovoltaic facilities $\mathcal{I}_i$, with $i\in \{1,\dots, 6\}$.
\item [Line 2:] We obtain all possible relative differences $\delta_{ik}(d^*)=x_{ik}$, for each pair of facilities $\mathcal{I}_i$ and $\mathcal{I}_k$. These results will be stored in a matrix as the one displayed in Table~\ref{tab}. Notice that the difference of each facility with itself is always equal to zero, as we can see in the main diagonal of the matrix.
\begin{table}[h!]
\begin{center}
\begin{tabular}{|c|cccccc|}
\hline
& $\mathcal {I}_1$ & $\mathcal {I}_2$  &  $\mathcal {I}_3$ &  $\mathcal {I}_4$ & $\mathcal {I}_5$ & $\mathcal {I}_6$\\ \hline
$\mathcal {I}_1$ & $0$  & $x_{12} $  &  $x_{13} $   &  $x_{14} $   & $x_{15} $   & $x_{16} $  \\  \hline
$\mathcal {I}_2$ & $x_{21} $  & $0$  &  $x_{23}$   &  $x_{24} $   & $x_{25} $   & $x_{26} $  \\  \hline
$\mathcal {I}_3$ & $x_{31} $  & $x_{32} $  &  $0$   &  $x_{34} $   & $x_{35} $   & $x_{36} $  \\  \hline
$\mathcal {I}_4$ & $x_{41} $  & $x_{42} $  &  $x_{43} $   &  $0$   & $x_{45} $   & $x_{46} $  \\  \hline
$\mathcal {I}_5$ & $x_{51} $  & $x_{52} $  &  $x_{53} $   &  $x_{54} $   & $0$   & $x_{56} $  \\  \hline
$\mathcal {I}_6$ & $x_{61} $ & $x_{62} $  &  $x_{63} $   &  $x_{64} $   & $x_{65} $   & $0$  \\  \hline
\end{tabular}
\caption{Matrix of relative differences of energy production for each pair of photovoltaic facilities in a specific day $d^*$.}\label{tab}
\end{center}
\end{table}

\item [Line 3:] We will evaluate the values collected in Table~\ref{tab} in their corresponding membership functions, that is, $\mu_{ik}$ will be applied to the value $x_{ik}$, for all $i,k\in\{1,\dots,6\}$ with $i\neq k$ (note that the main diagonal has no value due to it does not make sense to compare the performance of the installations with themselves). Once again we obtain a matrix collecting the resulting evaluations, as it is displayed in Table~\ref{tab2}. We recall that the meaning of these values indicate the degree in which the facilities are working properly. \begin{table}[h!]
\begin{center}
\begin{tabular}{|c|cccccc|}
\hline
& $\mathcal {I}_1$ & $\mathcal {I}_2$  &  $\mathcal {I}_3$ &  $\mathcal {I}_4$ & $\mathcal {I}_5$ & $\mathcal {I}_6$\\ \hline
$\mathcal {I}_1$ & $-$  & $\mu_{12}(x_{12}) $  &  $\mu_{13}(x_{13})$   &  $\mu_{14}(x_{14})$   & $\mu_{15}(x_{15})$   & $\mu_{16}(x_{16})$  \\  \hline
$\mathcal {I}_2$ & $\mu_{21}(x_{21})$  & $-$  &  $\mu_{23}(x_{23})$   &  $\mu_{24}(x_{24})$   & $\mu_{25}(x_{25})$   & $\mu_{26}(x_{26})$  \\  \hline
$\mathcal {I}_3$ & $\mu_{31}(x_{31})$  & $\mu_{32}(x_{32})$  &  $-$   &  $\mu_{34}(x_{34})$   & $\mu_{35}(x_{35})$   & $\mu_{36}(x_{36})$  \\  \hline
$\mathcal {I}_4$ & $\mu_{41}(x_{41})$  & $\mu_{42}(x_{42})$  &  $\mu_{43}(x_{43})$   &  $-$   & $\mu_{45}(x_{45})$   & $\mu_{46}(x_{46})$  \\  \hline
$\mathcal {I}_5$ & $\mu_{51}(x_{51})$  & $\mu_{52}(x_{52})$  &  $\mu_{53}(x_{53})$   &  $\mu_{54}(x_{54})$   & $-$   & $\mu_{56}(x_{56})$  \\  \hline
$\mathcal {I}_6$ & $\mu_{61}(x_{61})$ & $\mu_{62}(x_{62})$  &  $\mu_{63}(x_{63})$   &  $\mu_{64}(x_{64})$   & $\mu_{65}(x_{65})$   & $-$  \\  \hline
\end{tabular}
\caption{Matrix of evaluation of the relative differences in the corresponding membership function.}\label{tab2}
\end{center}
\end{table}

\item [Line 4:] In this step, we will interpret the results obtained in the matrix of Step 3. To interpret the results obtained for each facility, we will consider the rows of the matrix in Step 3 independently. For instance, if we are interested in analyzing the behavior of the facility $\mathcal {I}_i$, we will consider the $i$-th row of the matrix.  Notice that, the elements in the main diagonal will not be considered to carry out these computations and therefore, each vector is composed of 5 elements. 

 We will apply an OWA operator to the $i$-th row to obtain a single value that represents the global degree of proper performance for each facility. Therefore, the values
of this $i$-th row will be sorted from largest to smallest in a vector and they will be weighted according to the values of the vector $W=(0, \frac{1}{3},\frac{1}{3},\frac{1}{3}, 0)$, in order to assign the value 0 to the greatest and the least value of the ordered vector, with the goal of avoiding to consider extreme values since they are the most sensitive of being anomalous values. The obtained value $y_i$ will represent the global degree of suitable performance of the facility $\mathcal {I}_i$. {We have also tested   the weighted vector  $W=(0, \frac{1}{2},\frac{1}{2},0, 0)$, but it offers worse results, since less information is taken into account.} In the future, more cases of OWA operators will be studied. 

\item [Line 5:] Finally, the  value $y_i\in [0,1]$
 will be {empirically} identified with a linguistic label following the criteria presented below:\label{pag:labelsyi}

\begin{itemize}
\item If $y_i =1$, then $\mathcal {I}_i$  has a \emph{suitable performance (S)}.
\item If $y_i \in [{0.75}, 1)$,  then $\mathcal {I}_i$ has a \emph{lightly anomalous performance (LA)}.

\item If $y_i \in [{0.45}, {0.75})$ then $\mathcal {I}_i$ has an \emph{anomalous performance (A)}.
\item If $y_i \in (0, {0.45})$ then $\mathcal {I}_i$ has a \emph{very anomalous performance (VA)}.
\item If $y_i =0$ then $\mathcal {I}_i$  has a \emph{bad performance (B)}.
\end{itemize}

We will apply this procedure to all rows of the matrix of Step 3, obtaining a linguistic description about the behavior of each facility. 
\end{description}

In addition, the mechanism previously presented can be applied to a sequence of consecutive days, which provides a more detailed knowledge about the behavior of the facilities and consequently, a more robust support to make decisions. For example, with the sequential analysis, we can also detect if there is a gradual decrease in the performance of any facility.  
{To achieve this goal, a state machine \cite{Hopcroft2006} has been used. State machines allows us to describe the behavior of systems that are composed of a finite number of states. It also allows incorporating events that can occur in the states and that cause the evolution of the machine from a state to another state. For these reasons, they are suitable for simulating the behavior of each facility. A state machine is represented as a directed graph that is formally represented as $M = \{S, \Sigma, T, s_0\}$, where:} 
\begin{itemize}
	\item $S$  {is a set of finite states, each of which is represented by a circle.}	
	\item $\Sigma$ {is a set of input symbols.} 
	\item $T$ {is the set of transitions. Each transition is represented by an arrow labeled with the input symbols that causes a state change and that begins in one state and ends in another one, where both states can coincide.}
	\item $s_0$ {is the initial state and it is represented  by a double circle.} 
	\end{itemize}

{The possible states of each facility are the following:}
\begin{itemize}
	\item OK: The facility works properly.
	\item NRC: There is no reason to check the facility.
	\item SBC: The facility should be checked.
	\item KO: The facility does not work.
\end{itemize}

\begin{table}[h!]
	\begin{center}
		\begin{tabular}{|c|ccccc|}
			\hline
			      & $B$ & $VA$   & $A$   & $LA$   & $S$\\ \hline
			$OK$  & $KO$ & $SBC$ & $NRC$ & $NRC$ & $OK$ \\ 
			$NRC$ & $KO$ & $SBC$ & $SBC$ & $NRC$ & $OK$ \\ 
			$SBC$ & $KO$ & $KO$  & $SBC$ & $NRC$ & $OK$ \\ 
			$KO$  & $KO$ & $KO$  & $KO$  & $SBC$ & $NRC$ \\ \hline
		\end{tabular}
		\caption{Component $T$ of the states machine.}\label{tab:transitions}
	\end{center}
\end{table}

\begin{figure}[!ht]
	\centering
	\includegraphics[width=1.0\linewidth]{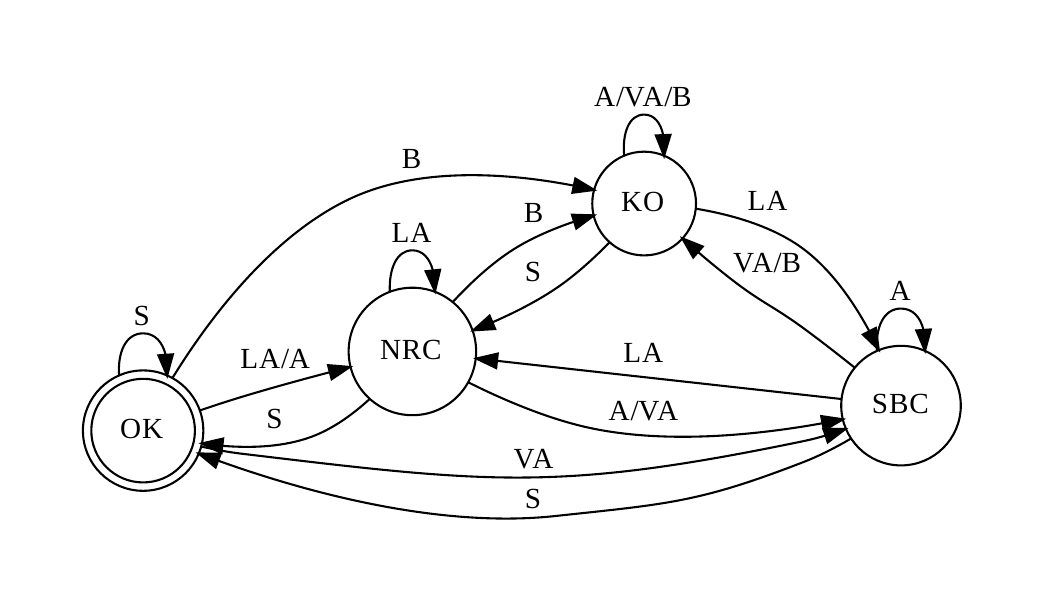}
	\caption{{{State} machine for every facility}}
	\label{fig:diagram}
\end{figure}

The designed machine is graphically displayed in Figure~\ref{fig:diagram} and it is formally represented as $M = \{S, \Sigma, T, s_0\}$, where $S=\{OK, NRC, SBC, KO\}$, $\Sigma=\{S, LA, A, VA, B\}$ (the linguistic labels defined for the value $y_i$ {in page~\pageref{pag:labelsyi}}), $T$ is given by Table \ref{tab:transitions} and $s_0 = \{OK\}$.

The nodes in Figure~\ref{fig:diagram} represent the {states} of the considered facility, whereas the edges indicate the performances of the considered facility in a specific day. Considering  as initial node the {state}  of a particular facility in a specific day, the transition to the next {state} for the following day will be carried out following the corresponding edge, that is, according to the performance of the facility in the following day. For example, if one given facility is ``OK'' as initial day ({state} for  Day 1) and  next day  the facility  has a lightly anomalous performance ``LA'' (performance of  Day 2), then the new {state} of the facility is ``NRC''({state} for   Day 2). Now, the current  {state} is ``NRC''. If the performance in the next day is ``B'' (performance of   Day 3), then  a transition to the {state} ``KO'' ({state} for   Day 3)   will be. However, if it has an anomalous performance ``A'' ({state} for  Day 3), the new {state} is ``SBC'' ({state} for   Day 3).

Thus, analyzing the performance in consecutive days, we can take decisions related to the {state} of the facilities.

\section{Illustrative example of the decision support system}\label{sec:example}

In this section, we will show the results obtained by the decision support system described in this paper. We have analyzed and contrasted the results obtained by the presented procedure  from January 1st to October 18th 2020, that is, 292 days. Nevertheless, since the automatic system returns the OWA value ($y_i$) and the {state} for each facility per each day, we have chosen some consecutive days in order to illustrate how the system may modify the decisions related to the {state} of  the facilities. In particular, we choose the period from  April 16th to 19th, 2020 and the facilities $\mathcal{I}_1$ and $\mathcal{I}_3$ to show how the system operates. 

The results obtained from the decision system for both facilities are shown in Table~\ref{ex:tab}. In this table, we can see that the facility $\mathcal{I}_1$ on April 16th had the {state} ``NRC'' (there is no reason to check the facility). Next day, the system returned $0$ as global degree of suitable performance ($y_1$), which means that the system identifies the value $y_1 = 0$ with the linguistic label \textit{bad performance} (B) and set the {state} of the facility $\mathcal{I}_1$ as ``KO'', according to the nodes of Figure~\ref{fig:diagram}. In this case, the facility does not work properly and should be inspected. Then, on April 18th the systems returned a global degree of $0.98$ and the system identified it as \textit{lightly anomalous performance} (LA) and it set the {state} to ``SBC'' (the facility should be checked), by the diagram in Figure~\ref{fig:diagram}. At this point, the company  examined the facility $\mathcal{I}_1$. On the last day chosen, we obtained the value $y_1 = 1$, that is, \textit{suitable performance} (S) and from the {state} ``SBC'' the system changed it to ``OK'' implying that the possible problem that arose has been already fixed. As we have seen, the facility $\mathcal{I}_1$ went through for all {states}. 
{This example shows  how the decision system automatically detects and alerts a problem (NRC and KO states), and as later, recovers OK once it has been solved, passing through the SBC state. Since the message of the state machine  for both detection and fix the problem can take more than one day, the system also return the value of the OWA operator in order to know the behaviour of each station during the day. Hence, the user have enough information for taking a decision. 
}

\begin{table}[h!]
	\centering
	\begin{tabular}{|c||c|c|c||c|c|c|}
		\hline
		Day	& $y_1$ & Transition & state $\mathcal{I}_1$ & $y_3$ & Transition & state $\mathcal{I}_3$ \\ \hline
		16-04-2020	& $0.88$ & LA & NRC & $1$ & S &OK \\ \hline
		17-04-2020	& $0$ & B & KO  & $1$ & S &OK\\ \hline
		18-04-2020	& $0.98$ & LA & SBC & $1$ & S &OK \\ \hline
		19-04-2020	& $1$ & S & OK & $1$ & S &OK\\ \hline
	\end{tabular}
	\caption{Results obtained by the automatic system for facilities $\mathcal{I}_1$ and $\mathcal{I}_3$.}
	\label{ex:tab}
\end{table}

On the other hand, we have analyzed all outputs obtained by the decision system by means of a confusion matrix~\cite{Berk2008} for each facility.
 We consider that our  system gives an alert when it returns either ``SBC'' or ``KO'' states. On the contrary,  no-alert  corresponds to ``NRC'' and ``OK'' states.

Table~\ref{confumatrix} shows the confusion matrices of each facility corresponding to our decision support system together with the obtained errors of the model, for all data collected in 2020. From the confusion matrices we can say that there are two different scenarios. On the one hand, we can mention that both facilities $\mathcal{I}_2$ and $\mathcal{I}_3$ have worked correctly during all the days of the considered period and our system {neither} report false positive {nor} false negative. On the other hand, we can observe that the rest of the facilities  have alerts and in this case, our system reports some false negatives.
The fact that the system does not detect all the alerts may be 
{given because of the noise  in the data. This fact will be improved in the future.}

\begin{table}[h!]
	\centering
	\begin{tabular}{|c||c|c||c|}
		\hline
		\textbf{$\mathcal{I}_1$} & No Alert & Alert &  Model error \\ \hline\hline
		No Alert 	 & 284      & 2 	& 	0.699\%	\\ \hline
		Alert		 & 0        & 6		& 	0\%	\\ \hline\hline
		Error of use & 0\%		& 25\% 	& 	0.685\%	\\ \hline
	\end{tabular}
	\vspace*{2ex}
	
	\begin{tabular}{|c||c|c||c|}
		\hline
		\textbf{$\mathcal{I}_2$} & No Alert & Alert &  Model error \\ \hline\hline
		No Alert 	 & 292      & 0 	& 	0\%	\\ \hline
		Alert		 & 0        & 0		& 	0\%	\\ \hline\hline
		Error of use & 0\%		& 0\% 	& 	0\%	\\ \hline
	\end{tabular}
	\vspace*{2ex}
	
	\begin{tabular}{|c||c|c||c|}
		\hline
		\textbf{$\mathcal{I}_3$} & No Alert & Alert &  Model error \\ \hline\hline
		No Alert 	 & 292      & 0 	& 	0\%	\\ \hline
		Alert		 & 0        & 0		& 	0\%	\\ \hline\hline
		Error of use & 0\%		& 0\% 	& 	0\%	\\ \hline
	\end{tabular}
	\vspace*{2ex}
	
	\begin{tabular}{|c||c|c||c|}
		\hline
		\textbf{$\mathcal{I}_4$} & No Alert & Alert &  Model error \\ \hline\hline
		No Alert 	 & 183      & 35 	& 	16.05\%	\\ \hline
		Alert		 & 0        & 74	& 	0\%	\\ \hline\hline
		Error of use & 0\%		& 32.11\%	& 	11.98\%	\\ \hline
	\end{tabular}
	\vspace*{2ex}
	
	\begin{tabular}{|c||c|c||c|}
		\hline
		\textbf{$\mathcal{I}_5$} & No Alert & Alert &  Model error \\ \hline\hline
		No Alert 	 & {68}       & 9 	&  {11.69}\%	\\ \hline
		Alert		 & 0        & {215}	& 	0\%	\\ \hline\hline
		Error of use & 0\%		& {4.02}\% 	& 	3.08\%	\\ \hline
	\end{tabular}
	\vspace*{2ex}
	
	\begin{tabular}{|c||c|c||c|}
		\hline
		\textbf{$\mathcal{I}_6$} & No Alert & Alert &  Model error \\ \hline\hline
		No Alert 	 & {184}       & 34 	& 	{15.60}\%	\\ \hline
		Alert		 & 0        & {74}	& 	0\%	\\ \hline\hline
		Error of use & 0\%		& {31.48}\%	& 	11.64\%	\\ \hline
	\end{tabular}
	
	\caption{Confusion matrices of each facility of the obtained results in 2020.}
	\label{confumatrix}
\end{table}

\begin{figure}[!h]
\centering
\begin{tikzpicture}
\begin{axis}[
 	legend columns=-1,
    legend style={at={(0.5,-0.15)},anchor=north},
	xlabel={Days},
	ylabel={$\rho_i/100$ per day},
	xmin=0, xmax=17,
	ymin=2, ymax=6.5,
	xtick={1,2,3,4,5,6,7,8,9,10,11,12,13,14,15,16,17},
	xticklabels={15,16,17,18,19,20,21,22,23,24,25,26,27,28,29,30}, 	
	ytick={2,2.5,3,3.5,4,4.5,5,5.5,6},
	ymajorgrids=true,
	grid style=dashed,
	]
	
	\addplot[
	color=black,
	mark=triangle,
	]
	coordinates {
		
		(1,5.23)(2,4.41)(3,4.77)(4,5.22)(5,5.22)(6,4.96)(7,5.04)(8,5.01)(9,4.91)(10,4.88)(11,4.93)(12,4.77)(13,4.88)(14,5.01)(15,4.90)(16,4.90)	
	};
	\addlegendentry{$\mathcal{I}_1$}
	
	\addplot[
	color=black,
	mark=square,
	]
	coordinates {
		(1,6.19)(2,6.08)(3,5.99)(4,6.13)(5,6.11)(6,5.93)(7,5.93)(8,5.95)(9,5.85)(10,5.81)(11,5.48)(12,5.67)(13,5.90)(14,5.81)(15,5.73)(16,5.89)	
	};
	\addlegendentry{$\mathcal{I}_2$}
	
	\addplot[
	color=black,
	mark=o,
	]
	coordinates {
		(1,5.50)(2,5.44)(3,5.34)(4,5.47)(5,5.41)(6,5.27)(7,5.28)(8,5.28)(9,5.11)(10,5.09)(11,5.08)(12,5.27)(13,5.27)(14,5.25)(15,5.12)(16,5.19)	
	};
	\addlegendentry{$\mathcal{I}_3$}
	
	\addplot[
	color=black,
	mark=diamond,
	]
	coordinates {
		(1,5.57)(2,5.49)(3,5.41)(4,5.56)(5,5.47)(6,4.63)(7,4.73)(8,4.68)(9,4.55)(10,4.63)(11,4.21)(12,4.53)(13,4.68)(14,4.58)(15,4.59)(16,4.73)	
	};
	\addlegendentry{$\mathcal{I}_4$}
	
	\addplot[
	color=black,
	mark=x,
	]
	coordinates {
		(1,2.44)(2,2.42)(3,2.33)(4,2.39)(5,2.44)(6,2.33)(7,2.35)(8,2.31)(9,2.23)(10,2.27)(11,2.27)(12,2.33)(13,2.33)(14,2.35)(15,2.25)(16,2.35)
	};
	\addlegendentry{$\mathcal{I}_5$}

	\addplot[
	color=black,
	mark=star,
	]
	coordinates {
		(1,4.15)(2,5.38)(3,5.29)(4,5.46)(5,4.09)(6,5.28)(7,4.00)(8,5.17)(9,3.81)(10,3.87)(11,4.92)(12,3.85)(13,5.22)(14,3.93)(15,5.02)(16,3.94)
		
	};
	\addlegendentry{$\mathcal{I}_6$}
\end{axis}
\end{tikzpicture}
\caption{Total performance of energy production of $\mathcal{I}_i$ in June 2020.}
\label{grafica_prod}
\end{figure}
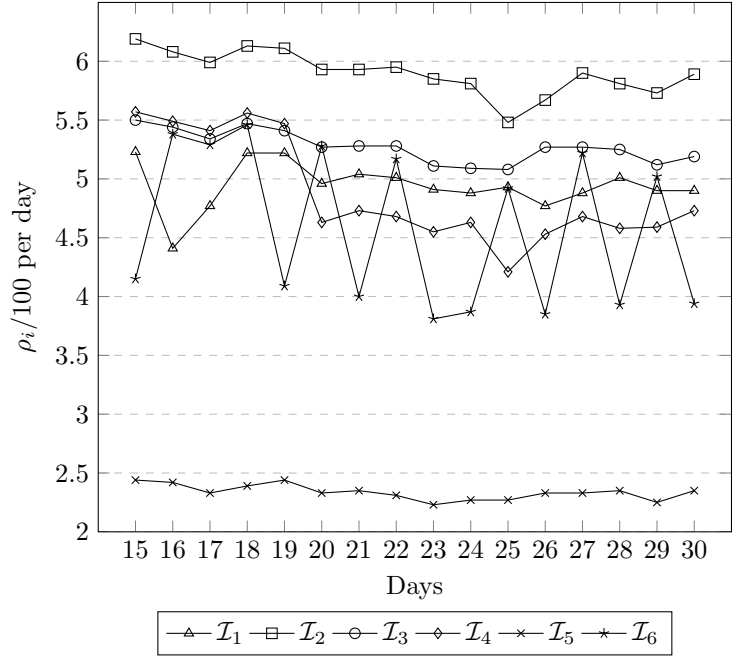

In particular, facilities $\mathcal{I}_1$ and $\mathcal{I}_5$ have  2 and 9 few false negatives, respectively, although the operation of $\mathcal{I}_1$ has been better than $\mathcal{I}_5$ taking into account the number of alerts of each one. For these two facilities the system {correctly classifies   the 
99,315\% and 96,92\% of the considered range of days, and} 
 detects 75\%  and 95.93\% of the alerts, respectively.  Similarly, for facilities $\mathcal{I}_4$ and $\mathcal{I}_6$ it {correctly classifies the    
88,02\% and 88,36\% of the days, and  detects} the 67.89\% and 69.1\% of the alerts, respectively. These two facilities have a significant number of alerts, but the nature of these alerts is different. For instance, the facility $\mathcal{I}_4$ had an inverter in bad conditions but the production in comparison with the rest of the facilities is not far enough for the system to detect the failure. 
However, the facility $\mathcal{I}_6$ had an irregular functioning for most of the time, having low production peaks and recovering the next day. As we previously explained, the system does not detect this type of scenario as an alert until the problem persists for a brief period of time, 
{but the user is alerted by the irregularity of the station.}
In addition, Figure~\ref{grafica_prod} illustrates these situations where the system does not immediately detect an alert. {For example, on day 16th, $\mathcal{I}_1$ has a low production peak but the system does not return an alert until day 17th because its recovery is not fast enough. After that, $\mathcal{I}_1$ returns to normality.} Moreover, we can observe on day 20th that $\mathcal{I}_4$ decreased its production and the system {alert with NRC. Until  June 25th, the system does not alert with SBC, where the differences with {the total performance of energy production of each facility} are more significant.} {Since NRC is not classified as alert in Table~\ref{confumatrix}, we are considering these days as FN although the system is sending a warning with the label NRC.}
 Lastly, the irregular performance that $\mathcal{I}_6$ had during these days can be perfectly appreciated in Figure~\ref{grafica_prod}. 
 
 {From the comments above on $\mathcal{I}_4$, we can see that the crisp character of the confusion matrix does not work perfectly to assess  the accuracy of our mechanism, transforming four different  states into only two values (true or false).}
 For example, it should be noted that most of the false negatives returned by the system correspond to the state ``NRC'', but this state can be interpreted as a warning of the start of a malfunction, {which} will be detected if the performance of the installation gets worse the next day. Specifically, all false negatives in $\mathcal{I}_1$, $\mathcal{I}_5$ and $\mathcal{I}_6$ correspond to the state ``NRC''. Moreover, in the facility $\mathcal{I}_4$, 21 of 35 false negatives also correspond to the state ``NRC''. Therefore, the system would go from having {80} false negatives to having only 14 false negatives, which corresponds to an error rate of only 4.79\%. 
In the future, we will try to improve the classification of the days (as correct and incorrect performances) and we also try to readjust the values of the intervals for a more optimal classification.

Notice that the system does not report any false positive, therefore all alerts reported are true and it prevents the company from making an effort to check the installation unnecessarily. In addition, according to Table~\ref{confumatrix}, we can ensure that the obtained errors are small and we can assume that our system is reliable.

\section{Conclusions and future work}\label{sec:conclusions} 

The main goal addressed in this paper has been to solve the problem of ``constant loss of energy production'', since when the loss of production is very small, it can be imperceptible from the user and so, it could not be detected during many months. To avoid this situation, a decision support system has been proposed for detecting anomalous behaviors in the energy production of the photovoltaic facilities over time.  For the development of this system,  different functions which describe the behaviors of the facilities has been designed, as well as a state machine that allows us to determine the state of each facility, taking into account the states and the performances obtained in the previous days for each facility. In addition, an experimental example has been carried out in order to test the proposed automatic decision system obtaining good results.

In the future, we will be focused on different improvements. {The first one will be to consider POWA  instead of OWA operators as aggregation operators and analyze if the use of these  operators improves the obtained results.} We also want to develop  the system in order to detect if the error in productivity is due to: dirty on the solar cells, investor in poor condition, investor has stopped working, {and} there are badly functioning {photovoltaic}  cells. {In addition, the mechanism will be extended with the consideration of meteorological variables, obtained from  weather stations installed in the facilities, in order to obtain a new mechanism to be available to be compared with the current ones that take into account these variables.}
Moreover, {in line with other failure prognosis approaches~\cite{kordestani21}},  we will carry out a predictive maintenance of each installation, anticipating in advance any deficiency that may happen in a particular installation.
Furthermore, a power generation predictability module will be designed. In this way, one could know approximately how much energy is going to be produced per installation. This fact can be very interesting if the production is needed to supply an own facility/building. For example, if the installation is going to heat the water of a sport center or school, and we know that not enough energy will be generated, we can take advantage of this information for performing some action in advance to heat the water. {These developments will be studied considering different mathematics and artificial intelligence tools~\cite{TFS:2020-acmr,fss:manlp2020,ChrisMedina14,moreno19,kou21,KOU2021109599,Medina2021:escim,7776941,9650769,YAO20221}.}


\begin{thebibliography}{10}

\bibitem{TFS:2020-acmr}
L.~{Antoni}, M.~E. {Cornejo}, J.~{Medina}, and E.~{Ramirez}.
\newblock Attribute classification and reduct computation in multi-adjoint
  concept lattices.
\newblock {\em IEEE Transactions on Fuzzy Systems}, pages 1--1, 2020.

\bibitem{ARISTONDO2016}
O.~Aristondo and M.~Ciommi.
\newblock The decompositions of rank-dependent poverty measures using ordered
  weighted averaging operators.
\newblock {\em International Journal of Approximate Reasoning}, 76:47 -- 62,
  2016.

\bibitem{Beliakov2020}
G.~Beliakov, S.~James, and J.-Z. Wu.
\newblock {\em Symmetric Fuzzy Measures: {OWA}}, pages 135--192.
\newblock Springer International Publishing, Cham, 2020.

\bibitem{BENAMMAR20201016}
R.~{Ben Ammar}, M.~{Ben Ammar}, and A.~Oualha.
\newblock Photovoltaic power forecast using empirical models and artificial
  intelligence approaches for water pumping systems.
\newblock {\em Renewable Energy}, 153:1016--1028, 2020.

\bibitem{Berk2008}
R.~A. Berk.
\newblock {\em Statistical learning from a regression perspective}.
\newblock Springer Series in Statistics. Springer Science+Business Media LLC,
  New York, NY, USA, 2008.

\bibitem{chang2014}
K.-H. Chang.
\newblock A decision support system for planning and coordination of hybrid
  renewable energy systems.
\newblock {\em Decision Support Systems}, 64:4 -- 13, 2014.

\bibitem{chenIVcharact2017}
Z.~Chen, L.~Wu, S.~Cheng, P.~Lin, Y.~Wu, and W.~Lin.
\newblock Intelligent fault diagnosis of photovoltaic arrays based on optimized
  kernel extreme learning machine and i-v characteristics.
\newblock {\em Applied Energy}, 204:912 -- 931, 2017.

\bibitem{CHIAO2019}
K.-P. Chiao.
\newblock Multiple criteria decision making for linguistic judgments with
  importance quantifier guided ordered weighted averaging operator.
\newblock {\em Information Sciences}, 474:48 -- 74, 2019.

\bibitem{fss:manlp2020}
M.~E. Cornejo, D.~Lobo, and J.~Medina.
\newblock Extended multi-adjoint logic programming.
\newblock {\em Fuzzy Sets and Systems}, 388:124--145, 2020.
\newblock Logic.

\bibitem{ChrisMedina14}
C.~Cornelis, J.~Medina, and N.~Verbiest.
\newblock Multi-adjoint fuzzy rough sets: Definition, properties and attribute
  selection.
\newblock {\em International Journal of Approximate Reasoning}, 55:412--426,
  2014.

\bibitem{demiguel2017}
L.~De~Miguel, D.~Paternain, I.~Lizasoain, G.~Ochoa, and H.~Bustince.
\newblock Some characterizations of lattice {OWA} operators.
\newblock {\em International Journal of Uncertainty, Fuzziness and
  Knowledge-Based Systems}, 25(Suppl. 1):5--17, 2017.

\bibitem{dhimishsixlayer2017}
M.~Dhimish, V.~Holmes, B.~Mehrdadi, and M.~Dales.
\newblock Diagnostic method for photovoltaic systems based on six layer
  detection algorithm.
\newblock {\em Electric Power Systems Research}, 151:26 -- 39, 2017.

\bibitem{dhimishPVCurves2017}
M.~Dhimish, V.~Holmes, B.~Mehrdadi, M.~Dales, and P.~Mather.
\newblock Photovoltaic fault detection algorithm based on theoretical curves
  modelling and fuzzy classification system.
\newblock {\em Energy}, 140:276 -- 290, 2017.

\bibitem{DURSO2019}
P.~D'Urso and J.~M. Leski.
\newblock Fuzzy clustering of fuzzy data based on robust loss functions and
  ordered weighted averaging.
\newblock {\em Fuzzy Sets and Systems}, 2019.

\bibitem{medina2013}
M.~El-Zekey, J.~Medina, and R.~Mesiar.
\newblock Lattice-based sums.
\newblock {\em Information Sciences}, 223:270 -- 284, 2013.

\bibitem{moreno19}
J.~M. Garc{\'{\i}}a, L.~Rodriguez{-}Benitez, L.~J. Linares, and
  G.~Trivi{\~{n}}o.
\newblock A linguistic extension of petri nets for the description of systems:
  An application to time series.
\newblock {\em {IEEE} Transactions on Fuzzy Systems}, 27(9):1818--1832, 2019.

\bibitem{garg2019}
H.~Garg.
\newblock Linguistic pythagorean fuzzy sets and its applications in
  multiattribute decision-making process.
\newblock {\em International Journal of Intelligent Systems}, 33(6):1234--1263,
  2018.

\bibitem{hajjiMultivariate2020}
M.~Hajji, M.-F. Harkat, A.~Kouadri, K.~Abodayeh, M.~Mansouri, H.~Nounou, and
  M.~Nounou.
\newblock Multivariate feature extraction based supervised machine learning for
  fault detection and diagnosis in photovoltaic systems.
\newblock {\em European Journal of Control}, 2020.
\newblock In press. https://doi.org/10.1016/j.ejcon.2020.03.004.

\bibitem{heldeweg2020}
M.~A. Heldeweg and S.~Saintier.
\newblock Renewable energy communities as `socio-legal institutions': A
  normative frame for energy decentralization?
\newblock {\em Renewable and Sustainable Energy Reviews}, 119:109518, 2020.

\bibitem{Hopcroft2006}
J.~Hopcroft, R.~Motwani, and J.~Ullman.
\newblock {\em Introduction to Automata Theory, Languages, and Computation}.
\newblock Pearson, 2006.

\bibitem{hussainANN2020}
M.~Hussain, M.~Dhimish, S.~Titarenko, and P.~Mather.
\newblock Artificial neural network based photovoltaic fault detection
  algorithm integrating two bi-directional input parameters.
\newblock {\em Renewable Energy}, 155:1272 -- 1292, 2020.

\bibitem{JAIME201464}
L.~G. Jaime, E.~E. Kerre, M.~Nachtegael, and H.~Bustince.
\newblock Consensus image method for unknown noise removal.
\newblock {\em Knowledge-Based Systems}, 70:64 -- 77, 2014.

\bibitem{review3}
L.~Jin and G.~Qian.
\newblock Owa generation function and some adjustment methods for owa operators
  with application.
\newblock {\em IEEE Transactions on Fuzzy Systems}, 24(1):168--178, 2016.

\bibitem{khalilComparative2020}
I.~U. {Khalil}, A.~{Ul-Haq}, Y.~{Mahmoud}, M.~{Jalal}, M.~{Aamir}, M.~U.
  {Ahsan}, and K.~{Mehmood}.
\newblock Comparative analysis of photovoltaic faults and performance
  evaluation of its detection techniques.
\newblock {\em IEEE Access}, 8:26676--26700, 2020.

\bibitem{review1}
A.~Kishor, A.~K. Singh, S.~Sonam, and N.~R. Pal.
\newblock A new family of owa operators featuring constant orness.
\newblock {\em IEEE Transactions on Fuzzy Systems}, 28(9):2263--2269, 2020.

\bibitem{kobylinski2020}
P.~Kobylinski, M.~Wierzbowski, and K.~Piotrowski.
\newblock High-resolution net load forecasting for micro-neighbourhoods with
  high penetration of renewable energy sources.
\newblock {\em International Journal of Electrical Power \& Energy Systems},
  117:105635, 2020.

\bibitem{review2}
M.~Kordestani, A.~Chibakhsh, and M.~Saif.
\newblock A control oriented cyber-secure strategy based on multiple sensor
  fusion.
\newblock In {\em 2019 IEEE International Conference on Systems, Man and
  Cybernetics (SMC)}, pages 1875--1881, 2019.

\bibitem{kordestani18}
M.~Kordestani, A.~Mirzaee, A.~A. Safavi, and M.~Saif.
\newblock Maximum power point tracker (mppt) for photovoltaic power systems-a
  systematic literature review.
\newblock In {\em 2018 European Control Conference (ECC)}, pages 40--45, 2018.

\bibitem{kordestani21}
M.~Kordestani, M.~Saif, M.~E. Orchard, R.~Razavi-Far, and K.~Khorasani.
\newblock Failure prognosis and applications: A survey of recent literature.
\newblock {\em IEEE Transactions on Reliability}, 70(2):728--748, 2021.

\bibitem{review4}
M.~Kordestani, M.~F. Samadi, M.~Saif, and K.~Khorasani.
\newblock A new fault diagnosis of multifunctional spoiler system using
  integrated artificial neural network and discrete wavelet transform methods.
\newblock {\em IEEE Sensors Journal}, 18(12):4990--5001, 2018.

\bibitem{kou21}
G.~Kou, {\"O}.~{Olgu Akdeniz}, H.~Dinçer, and S.~Y\"uksel.
\newblock Fintech investments in european banks: a hybrid it2 fuzzy
  multidimensional decision-making approach.
\newblock {\em Financial Innovation}, 7:39, 2021.

\bibitem{KOU2021109599}
G.~Kou, H.~Xiao, M.~Cao, and L.~H. Lee.
\newblock Optimal computing budget allocation for the vector evaluated genetic
  algorithm in multi-objective simulation optimization.
\newblock {\em Automatica}, 129:109599, 2021.

\bibitem{leszczyna2019}
R.~Leszczyna, T.~Wallis, and M.~R. Wr\'obel.
\newblock Developing novel solutions to realise the european energy --
  information sharing \& analysis centre.
\newblock {\em Decision Support Systems}, 122:113067, 2019.

\bibitem{7776941}
G.~Li, G.~Kou, and Y.~Peng.
\newblock A group decision making model for integrating heterogeneous
  information.
\newblock {\em IEEE Transactions on Systems, Man, and Cybernetics: Systems},
  48(6):982--992, 2018.

\bibitem{xia2019}
X.~Liu, Y.~Xu, R.~Montes, Y.~Dong, and F.~Herrera.
\newblock Analysis of self-confidence indices-based additive consistency for
  fuzzy preference relations with self-confidence and its application in group
  decision making.
\newblock {\em International Journal of Intelligent Systems}, 34(5):920--946,
  2019.

\bibitem{LIZASOAIN2013}
I.~Lizasoain and C.~Moreno.
\newblock {OWA} operators defined on complete lattices.
\newblock {\em Fuzzy Sets and Systems}, 224:36 -- 52, 2013.
\newblock Theme: Aggregation functions and implications.

\bibitem{manohar2020}
M.~Manohar, E.~Koley, and S.~Ghosh.
\newblock Microgrid protection under weather uncertainty using joint
  probabilistic modeling of solar irradiance and wind speed.
\newblock {\em Computers \& Electrical Engineering}, 86:106684, 2020.

\bibitem{marin2019}
L.~G. Mar\'in, N.~Cruz, D.~S\'aez, M.~Sumner, and A.~Nuñez.
\newblock Prediction interval methodology based on fuzzy numbers and its
  extension to fuzzy systems and neural networks.
\newblock {\em Expert Systems with Applications}, 119:128 -- 141, 2019.

\bibitem{mattiussi2014}
A.~Mattiussi, M.~Rosano, and P.~Simeoni.
\newblock A decision support system for sustainable energy supply combining
  multi-objective and multi-attribute analysis: An australian case study.
\newblock {\em Decision Support Systems}, 57:150 -- 159, 2014.

\bibitem{Medina12Charac}
J.~Medina.
\newblock Characterizing when an ordinal sum of t-norms is a t-norm on bounded
  lattices.
\newblock {\em Fuzzy Sets and Systems}, 202:75--88, 2012.

\bibitem{Medina2021:escim}
J.~Medina, J.~Moreno-García, E.~Ramírez-Poussa, and L.~T. Kóczy.
\newblock Mathematics and computational intelligence synergies for emerging
  challenges.
\newblock {\em International Journal of Computational Intelligence Systems},
  14:818--820, 2021.

\bibitem{MEDINA202138}
J.~Medina and R.~R. Yager.
\newblock {OWA} operators with functional weights.
\newblock {\em Fuzzy Sets and Systems}, 414:38--56, 2021.
\newblock Aggregation Functions.

\bibitem{review5}
R.~Mesiar, F.~Kouchakinejad, A.~\v~Sipo\v~sov\'a, and M.~Mashinchi.
\newblock Owa operators on complete lattices.
\newblock {\em IEEE Transactions on Fuzzy Systems}, 26(6):3884--3887, 2018.

\bibitem{morato2020}
M.~M. Morato, P.~R. Mendes, J.~E. Normey-Rico, and C.~Bordons.
\newblock {LPV}-{MPC} fault-tolerant energy management strategy for renewable
  microgrids.
\newblock {\em International Journal of Electrical Power \& Energy Systems},
  117:105644, 2020.

\bibitem{review6}
M.~Mousavi, M.~Moradi, A.~Chaibakhsh, M.~Kordestani, and M.~Saif.
\newblock Ensemble-based fault detection and isolation of an industrial gas
  turbine.
\newblock In {\em 2020 IEEE International Conference on Systems, Man, and
  Cybernetics (SMC)}, pages 2351--2358, 2020.

\bibitem{PU201924}
X.~Pu, L.~Jin, R.~Mesiar, and R.~R. Yager.
\newblock Continuous parameterized families of rim quantifiers and
  quasi-preference with some properties.
\newblock {\em Information Sciences}, 481:24 -- 32, 2019.

\bibitem{rodriguesMLPV2020}
S.~Rodrigues, G.~M{\"u}tter, H.~G. Ramos, and F.~Morgado-Dias.
\newblock Machine learning photovoltaic string analyzer.
\newblock {\em Entropy}, 22(2):205, 2020.

\bibitem{SAID20201274}
M.~I. Said, M.~Steiner, G.~Siefer, and A.~H. Arab.
\newblock Maximum power output prediction of {HCPV FLATCON}® module using an
  {ANN} approach.
\newblock {\em Renewable Energy}, 152:1274 -- 1283, 2020.

\bibitem{samara2020}
S.~Samara and E.~Natsheh.
\newblock Intelligent pv panels fault diagnosis method based on {NARX} network
  and linguistic fuzzy rule-based systems.
\newblock {\em Sustainability}, 12(5):2011, 2020.

\bibitem{mesiar08}
S.~Saminger-Platz, E.~Klement, and R.~Mesiar.
\newblock On extension of triangular norms on bounded lattices.
\newblock {\em Indagationes Mathematicae}, 19(1):135--150, 2008.

\bibitem{sandeep2019}
S.~R. Sandeep and R.~Nandihalli.
\newblock Optimal sizing in hybrid renewable energy system with the aid of
  opposition based social spider optimization.
\newblock {\em Journal of Electrical Engineering {\&} Technology}, 15:433--440,
  2020.

\bibitem{SIMINSKI2017}
K.~Siminski.
\newblock Fuzzy weighted c-ordered means clustering algorithm.
\newblock {\em Fuzzy Sets and Systems}, 318:1 -- 33, 2017.
\newblock Theme: Clustering and Image Processing.

\bibitem{singh2019}
R.~{Singh} and R.~C. {Bansal}.
\newblock Optimization of an autonomous hybrid renewable energy system using
  reformed electric system cascade analysis.
\newblock {\em IEEE Transactions on Industrial Informatics}, 15(1):399--409,
  2019.

\bibitem{takashima2009}
T.~Takashima, J.~Yamaguchi, K.~Otani, T.~Oozeki, K.~Kato, and M.~Ishida.
\newblock Experimental studies of fault location in pv module strings.
\newblock {\em Solar Energy Materials and Solar Cells}, 93(6):1079--1082, 2009.
\newblock 17th International Photovoltaic Science and Engineering Conference.

\bibitem{torreglosa2016}
J.~P. Torreglosa, P.~Garc\'ia-Triviño, L.~M. Fern\'andez-Ram\'irez, and
  F.~Jurado.
\newblock Control based on techno-economic optimization of renewable hybrid
  energy system for stand-alone applications.
\newblock {\em Expert Systems with Applications}, 51:59 -- 75, 2016.

\bibitem{9650769}
H.~Xiao, Y.~Yan, G.~Kou, and S.~Wu.
\newblock Optimal inspection policy for a single-unit system considering two
  failure modes and production wait time.
\newblock {\em IEEE Transactions on Reliability}, pages 1--13, 2021.

\bibitem{yager88}
R.~Yager.
\newblock On ordered weighted averaging aggregation operators in multi-criteria
  decision making.
\newblock {\em IEEE Transactions on Systems, Man and Cybernetics}, 18:183--190,
  1988.

\bibitem{YAGER201788}
R.~R. Yager.
\newblock {OWA} aggregation of multi-criteria with mixed uncertain
  satisfactions.
\newblock {\em Information Sciences}, 417:88 -- 95, 2017.

\bibitem{yagerkacprzyk97}
R.~R. Yager and J.~Kacprzyk.
\newblock {\em The Ordered Weighted Averaging Operators: Theory and
  Applications}.
\newblock Kluwer: Norwell, MA, 1997.

\bibitem{YAO20221}
J.~Yao, J.~Medina, Y.~Zhang, and D.~\'{S}l\c{e}zak.
\newblock Formal concept analysis, rough sets, and three-way decisions.
\newblock {\em International Journal of Approximate Reasoning}, 140:1--6, 2022.

\bibitem{Z1}
L.~A. Zadeh.
\newblock Fuzzy sets.
\newblock {\em Information and Control}, 8:338--353, 1965.

\end{thebibliography}

\end{document}